\documentclass{aa}  \usepackage{natbib}
\bibpunct{(}{)}{;}{a}{}{,} \usepackage{graphicx}
\usepackage{booktabs}
\usepackage{txfonts}
\usepackage[colorlinks=true, allcolors=blue]{hyperref}
\usepackage{xcolor}

\newcommand{\carrlong}{$\Phi$}
\newcommand{\carrlat}{$\Theta$}

\begin{document}

   \title{Multi-spacecraft study of the solar wind at solar minimum: dependence on latitude and transient outflows}

   \author{R. Laker\inst{1}\thanks{Corresponding author: Ronan Laker \email{ronan.laker15$@$imperial.ac.uk}}
   \and
   T. S. Horbury\inst{1}
   \and
   S. D. Bale\inst{1,2,3}
   \and
   L. Matteini\inst{1}
   \and
   T. Woolley\inst{1}
   \and
   L. D. Woodham\inst{1}
   \and
   J. E. Stawarz\inst{1}
   \and
   E. E. Davies\inst{1}
   \and
   J. P. Eastwood\inst{1}
   \and
   M. J. Owens\inst{4}
   \and
   H. O'Brien\inst{1}
   \and
   V. Evans\inst{1}
   \and
   V. Angelini\inst{1}
   \and
   I. Richter\inst{5}
   \and
   D. Heyner\inst{5}
   \and
   C. J. Owen\inst{6}
   \and
   P. Louarn\inst{7}
   \and
   A. Federov\inst{7}
   }

   \institute{Imperial College London, South Kensington Campus, London, SW7 2AZ, UK
   \and
   Physics Department, University of California, Berkeley, CA 94720-7300, USA
   \and
   Space Sciences Laboratory, University of California, Berkeley, CA 94720-7450, USA
   \and
   Department of Meteorology, University of Reading, Earley Gate, PO Box 243, RG6 6BB Reading, UK
   \and
   Technical University of Braunschweig, Braunschweig, Germany
   \and
   Mullard Space Science Laboratory, University College London, Holmbury St. Mary, Dorking, Surrey RH5 6NT, UK
   \and
   Institut de Recherche en Astrophysique et Planétologie, 9, Avenue du Colonel ROCHE, BP 4346, 31028 Toulouse Cedex 4, France
   }

   \date{Received XXXX; accepted YYYY}

  \abstract
{The recent launches of Parker Solar Probe (PSP), Solar Orbiter (SO) and BepiColombo, along with several older spacecraft, have provided the opportunity to study the solar wind at multiple latitudes and distances from the Sun simultaneously.}
{We take advantage of this unique spacecraft constellation, along with low solar activity across two solar rotations between May and July 2020, to investigate how the solar wind structure, including the Heliospheric Current Sheet (HCS), varies with latitude.}
{We visualise the sector structure of the inner heliosphere by ballistically mapping the polarity and solar wind speed from several spacecraft onto the Sun's source surface. We then assess the HCS morphology and orientation with the in situ data and compare with a predicted HCS shape.}
{We resolve ripples in the HCS on scales of a few degrees in longitude and latitude, finding that the local orientation of sector boundaries were broadly consistent with the shape of the HCS but were steepened with respect to a modelled HCS at the Sun. We investigate how several CIRs varied with latitude, finding evidence for the compression region affecting slow solar wind outside the latitude extent of the faster stream. We also identified several transient structures associated with HCS crossings, and speculate that one such transient may have disrupted the local HCS orientation up to five days after its passage.}
{We have shown that the solar wind structure varies significantly with latitude, with this constellation providing context for solar wind measurements that would not be possible with a single spacecraft. These measurements provide an accurate representation of the solar wind within $\pm 10^{\circ}$ latitude, which could be used as a more rigorous constraint on solar wind models and space weather predictions. In the future, this range of latitudes will increase as SO's orbit becomes more inclined.}


   \maketitle

\section{Introduction}
Early spacecraft measurements in the ecliptic plane revealed that although the magnetic field was aligned with the Parker spiral \citep{Parker1958}, it would reverse direction several times per solar rotation, either pointed away (positive polarity) or towards (negative polarity) the Sun \citep{Wilcox1965}.
This, along with a single polarity effect observed out of the ecliptic plane \citep{Rosenberg1969}, led to the idea of a warped Heliospheric Current Sheet (HCS), that extends throughout the heliosphere and separates opposing magnetic polarities \citep{Alfven1977, Smith2001b}.

At solar minimum, when the Sun's magnetic field can be well described by a dipole configuration, the HCS has a limited latitudinal extent, due to its relation to the tips of closed field lines in the equatorial streamer belt \citep{Gosling1981, Hoeksema1983}.
The HCS also exhibits a low local inclination, meaning that all parts of the HCS are relatively parallel with the solar equator \citep{Smith1986, Peng2017}.
Therefore, at solar minimum the HCS is generally flat with a well defined shape that can persist for several solar rotations \citep{Thomas1981, Riley2002}.
However, at solar maximum, where the magnetic field at the poles is no longer dominant, the HCS covers a wider range of latitudes accompanied by a higher local inclination \citep{Suess1993, Owens2012}.
This results in a much more complex HCS structure, where there is no longer a single polarity per hemisphere \citep{Hoeksema1991}.

The location and shape of the HCS has a direct impact on the sector polarity and solar wind conditions seen by a spacecraft in the solar wind.
Therefore, by modelling the location of the HCS, contextual information about the source region of the solar wind can be gained.
This can be achieved by tracking the location of the streamer belt in white light images \citep{Wang1997,Robbrecht2012, Rouillard2020a}, or by using a numerical model driven by remote sensing observations \citep{Odstrcil2003}.
The widely adopted potential field source surface (PFSS) model assumes a current-free corona and radial solar wind expansion past a spherical source surface \citep{Schatten1969, Altschuler1969}.
Despite its simplicity, PFSS models have been shown to compare well with more sophisticated numerical models \citep{Riley2006}, as well as in situ measurements at 1 AU \citep{Jian2015} and closer to the Sun \citep{Badman2020a, Panasenco2020}.
The local inclination of the HCS predicted by the PFSS model has also been shown to be consistent with in situ measurements except in those cases where transients, such as coronal mass ejections (CMEs), are present \citep{Klein1980, Burton1994, Peng2017}.
In situ manifestations of CMEs are commonly observed around HCS crossings where it has been argued that they carry the sector boundary, rather than being structures that drape or push the HCS aside \citep{Gosling1987, Crooker1993, Crooker1996, forsyth1997}.
While transients can disrupt the local HCS structure, it is generally accepted that the HCS reforms to its original state, although the timescale of this process is debated \citep{Zhao1996, Crooker1998, Blanco2011}.

During solar minimum, the solar wind mirrors the magnetic structure of the Sun, with fast solar wind (> 600 km s$^{-1}$) emanating from regions of open field at the Sun's poles, known as coronal holes \citep[CHs, ][]{Mccomas1998}, and a slower, more variable, solar wind surrounding the HCS at low latitudes \citep{Zhao1981, Gosling1981, Pizzo1994}.
As a result of coronal structure, solar wind of varying speeds can exist at the same heliographic latitude, which can create a co-rotating interaction region (CIR), providing that the solar wind sources are time stationary \citep{Smith1976}.
A typical CIR consists of a rarefaction at the trailing edge of the fast stream, and a compression region at the leading edge, which envelops the HCS as the CIR develops \citep{Gosling1999}.
CIRs also have a distinctive east-west flow deflection across the stream boundary due to the Sun's rotation \citep{Richardson2018b}, although the meridional flow deflections depend on the CIR tilt \citep{Siscoe1969}.
This effect has been observed at latitudes $> 30^{\circ}$ with Ulysses \citep{Gosling1993}, and have been shown to significantly affect the structure of the HCS \citep{Pizzo1994, Lee2000}.
The compression at the leading edge of a CIR can produce a planar magnetic structure \citep[PMS, ][]{Nakagawa1989}, where the local magnetic field is forced to lie in the same plane as the stream interface \citep{Broiles2012}.
If these magnetic field deflections have a significant southward component then this, along with increased density and speed, can drive space weather effects at Earth \citep{Tsurutani2006}.
Therefore, it is important to understand how CIRs vary with both distance and latitude.

Due to the restriction of single point measurements, many studies have relied on large statistics to investigate CIR properties and development \citep{Richter1986, Jian2006}.
However, with the recent launches of Solar Orbiter \citep[SO, ][]{Muller2013}, Parker Solar Probe \citep[PSP, ][]{Fox2016} and BepiColombo \citep{Benkhoff2010}, there are now an unprecedented number of active spacecraft in the inner heliosphere.
These, along with the other missions such as Wind \citep{Ogilvie1997} and the Solar Terrestrial Relations Observatory \citep[STEREO, ][]{Kaiser2008a}, provide a constellation of spacecraft that can be used collectively to improve upon single spacecraft measurements.
Recently, several studies have investigated how certain solar wind features evolve from the close proximity PSP measurements out to 1 AU \citep{Szabo2020, Panasenco2020, Allen2021}.
In this paper, we demonstrate that this spacecraft configuration can be used to investigate how individual features in the solar wind vary with latitude and distance from the Sun.
To visualise the solar wind's sector structure, we ballistically map in situ data from the available spacecraft onto the Sun's source surface, as outlined in Section \ref{sec:method}.
We then empirically determine the HCS shape, and compare it to a PFSS model in Section \ref{sec:results:hcs}, while also demonstrating that this technique can identify coherent structures measured by several spacecraft at a range of latitudes (Sections \ref{sec:results:cir} and \ref{sec:results:iden}).
In Section \ref{sec:results:transient}, we present observations of several transient structures and examine their effect on the associated sector boundaries.
Finally, our conclusions are presented in Section \ref{sec:conclusions}.
 
\section{Methods}\label{sec:method}

In this paper, we consider two solar rotations with low solar activity, CR2231 and CR2232, which span from 21 May to 15 July 2020.
We refer to the position of the spacecraft in Carrington coordinates, specifically the `IAU\_SUN' frame in NAIF's SPICE framework \citep{Acton2018}.
This frame rotates with the Sun's surface, at a rate of $14.18^{\circ}$/day \citep{Archinal2011}, allowing each point on the Sun to be described by a longitude, \carrlong, and latitude, \carrlat, which can be extended into the heliosphere by including the distance from the Sun's surface, $R$.
We note that the solar equator is not aligned with the ecliptic plane, meaning that the Carrington latitude of a spacecraft orbiting in the ecliptic plane will depend on its Carrington longitude.

\begin{table}[ht!]
\resizebox{\hsize}{!}{\begin{tabular}{@{}llll@{}}
\toprule
Spacecraft                       & Instrument                   & Type                             & Reference              \\ \midrule
\multicolumn{1}{l|}{PSP}         & \multicolumn{1}{l|}{FIELDS}  & \multicolumn{1}{l|}{Mag}         & \citet{Bale2016}       \\
\multicolumn{1}{l|}{PSP}         & \multicolumn{1}{l|}{SPC}     & \multicolumn{1}{l|}{Plasma}      & \citet{Kasper2016},    \\
\multicolumn{1}{l|}{}            & \multicolumn{1}{l|}{}        & \multicolumn{1}{l|}{}            & \citet{Case2020}       \\
\multicolumn{1}{l|}{PSP}         & \multicolumn{1}{l|}{SPAN-e}  & \multicolumn{1}{l|}{$e^{-}$ PAD} & \citet{Kasper2016},    \\
\multicolumn{1}{l|}{}            & \multicolumn{1}{l|}{}        & \multicolumn{1}{l|}{}            & \citet{Whittlesey2020} \\
\multicolumn{1}{l|}{SO}        & \multicolumn{1}{l|}{MAG}     & \multicolumn{1}{l|}{Mag}         & \citet{Horbury2020b}   \\
\multicolumn{1}{l|}{SO}        & \multicolumn{1}{l|}{PAS}     & \multicolumn{1}{l|}{Plasma}      & \citet{Owen2020}       \\
\multicolumn{1}{l|}{BepiColombo} & \multicolumn{1}{l|}{MAG}     & \multicolumn{1}{l|}{Mag}         & \citet{Glassmeier2010} \\
\multicolumn{1}{l|}{}            & \multicolumn{1}{l|}{}        & \multicolumn{1}{l|}{}            & \citet{Heyner2020}     \\
\multicolumn{1}{l|}{STEREO-A}    & \multicolumn{1}{l|}{IMPACT}  & \multicolumn{1}{l|}{Mag}         & \citet{Acuna2008}      \\
\multicolumn{1}{l|}{STEREO-A}    & \multicolumn{1}{l|}{PLASTIC} & \multicolumn{1}{l|}{Plasma}      & \citet{Galvin2008}     \\
\multicolumn{1}{l|}{Wind}        & \multicolumn{1}{l|}{MFI}     & \multicolumn{1}{l|}{Mag}         & \citet{Lepping1995}    \\
\multicolumn{1}{l|}{Wind}        & \multicolumn{1}{l|}{SWE}     & \multicolumn{1}{l|}{Plasma}      & \citet{Ogilvie1995}    \\
\multicolumn{1}{l|}{Wind}        & \multicolumn{1}{l|}{3DP}     & \multicolumn{1}{l|}{$e^{-}$ PAD} & \citet{Lin1995a}       \\ \bottomrule
\end{tabular}}
\caption{List of the different measurements used in this paper, for the period spanning 21 May to 15 July 2020. All spacecraft had magnetic field (Mag) data available, although only Wind and PSP had bulk plasma parameters and electron pitch angle distributions (PADs).}
\label{table:data}
\end{table}

We used data from a wide variety of spacecraft throughout the inner heliosphere, where a full list of each available dataset can be found in Table \ref{table:data}.
PSP reached a perihelion of $0.13$ AU on the 7th June 2020 during its fifth solar encounter and continued taking data out to $\sim0.5$ AU.
As seen in Table \ref{table:data}, both PSP and Wind provided all three types of data used in this paper: magnetic field, bulk proton parameters and the electron strahl.
This was not the case for SO, which was launched in February 2020, and was in the commissioning phase during this paper's period of interest.
Therefore, there are only a few days of bulk plasma data from the Proton-Alpha Sensor (PAS), in early June when SO was predicted to encounter the tail of comet ATLAS \citep{Jones2020}.
However, the MAG instrument continuously measured the magnetic field throughout these two solar rotations, at distances ranging from $0.51$ AU at perihelion to $0.63$ AU.

The BepiColombo spacecraft was in the cruise phase ahead of orbiting Mercury at the end of 2025 \citep{Steiger2020}, and had completed an Earth flyby on the 10 April 2020.
Therefore, BepiColombo was close to Earth during this period, with a radial distance from the Sun decreasing from $0.99$ AU to $0.85$ AU.
The magnetometer (MAG) aboard the Mercury Planetary Orbiter (MPO) was operating at 16 vector/second.
This data has been ground calibrated regarding temperature dependence of sensitivity, misalignment and sensor offset.
Furthermore, a quasi-static spacecraft disturbance field, derived from long term in-flight experience, is removed, but any time dependent disturbances are still visible in the data.
However, we mitigated this issue by assessing the data for artificial signals and only considering an average magnetic field over several hours to investigate the polarity and relative magnetic field strength.

We estimated the shape of the HCS and the distribution of open field lines by implementing the PFSS model using the open source \textit{pfsspy} Python package \citep{Yeates2018, pfsspy}.
We supplied a synoptic magnetogram from the Air Force Data Assimilative Photospheric Flux Transport (ADAPT) model, which attempts to forecast flux transport, allowing for more accurate results outside of the Earth's field of view \citep{Hickman2015}.
The underlying assumptions of the PFSS model are that the corona is current free and the field becomes radial past some arbitrary source surface.
Recently, several papers have suggested using a variable source surface height \citep{Badman2020a, Panasenco2020}.
However, we opted for a constant source surface height so that all spacecraft measurements could be mapped back to a single reference distance.
We use an average of the 12 realisations from a single ADAPT model, with a constant source surface height of $2$ solar radii, $R_{\bigodot}$, which has been shown to provide a better match to the magnetic field polarity than the widely used value of $2.5 R_{\bigodot}$ \citep{Nikolic2019, Badman2020a}.

We performed this mapping using a simple ballistic propagation \citep{Nolte1973, Stansby2019}, which assumes a constant radial solar wind speed, $V_{SW}$.
This allows us to compare spacecraft measurements taken at different distances, as well as with the PFSS model.
This technique transforms the spacecraft's position at a time $t$, described by $\Phi_{SC}$ and $R_{SC}$, to a longitude at the source surface given by:
\begin{equation}
    \Phi_{SS}(t) = \Phi_{SC}(t) + \frac{\Omega(R_{SC}(t) - 2R_{\bigodot}) }{V_{SW}(t)},
\end{equation}
where $\Omega$ is the solar rotation rate used by the IAU\_SUN frame, and $_{SC}$ denotes that a property belongs to a spacecraft.
Throughout this paper we refer to $\Phi_{SS}$ as longitude, unless otherwise specified.
We used a six-hour average to determine $V_{SW}$, where bulk proton data was available (Table \ref{table:data}).
There is no change to the latitude as a radial flow is assumed.
Since BepiColombo orbited close to Earth we used data from Wind to provide a contextual solar wind speed to be used for the mapping.
We assumed a speed of $350$ km s$^{-1}$ for the SO spacecraft, based on Wind observations across the two solar rotations.
A deviation of $50$ km s$^{-1}$ from this assumed speed would have resulted in a $\sim5^{\circ}$ error in $\Phi_{SS}$ for SO at $\sim0.5$ AU.
There is also an inherent uncertainty in this ballistic mapping owing to the interaction between different solar wind parcels, that can only be estimated by a more sophisticated model than used here \citep[e.g.][]{Roussev2003, Owens2020}.

In an effort to establish the sector structure of the solar wind, we determined the magnetic polarity of each six-hour period used in the mapping.
Under normal solar wind conditions, the interplanetary magnetic field lies along the Parker spiral \citep{Parker1958}, either pointing towards ($\phi_{PS, T}$) or away ($\phi_{PS, A}$) from the Sun.
We calculated the Parker spiral angle for each six-hour periods, using the same $R_{SC}$, $V_{SW}$ and $\Omega$ values from the ballistic mapping.
We refer to the magnetic field in Radial-Tangential-Normal (RTN) coordinates, where \vec{R} points from the Sun to the spacecraft, \vec{N} is the component of the solar north direction perpendicular to \vec{R}, and \vec{T} completes the right-handed set.
We express the magnetic field direction as angles in the R-T ($\phi$), and T-N planes ($\theta$), where $\phi = 0^{\circ}, \ \theta = 0^{\circ}$ is along \vec{R} and $\phi = 90^{\circ}, \ \theta = 0^{\circ}$ is parallel to \vec{T}.
We assigned the magnetic polarity as being outwards when $\phi_{PS, A}- 45^{\circ} < \phi < \phi_{PS, A}+ 45^{\circ}$, and inwards when $\phi_{PS, T}- 45^{\circ} < \phi < \phi_{PS, T}+ 45^{\circ}$. If $\phi$ lay outside this range of angles then we assigned the polarity as mixed.

While the magnetic field data allowed us to observe when the magnetic polarity changed, as is expected at a HCS crossing, it does not necessarily mark the location of the sector boundary \citep{Crooker2010, Owens2013}.
Therefore, to investigate the true connectivity of our identified events, we studied the pitch angle distribution (PAD) of the suprathermal electron population, where the pitch angle ranged from $0^{\circ}$ (parallel) to $180^{\circ}$ (anti-parallel) to the local magnetic field direction \citep{Feldman1975, Pilipp1987}.
This population, called strahl, is created in the solar corona \citep{Rosenbauer1977}, so it is expected that the interplanetary magnetic field with (inwards) outwards magnetic polarity will exhibit a (anti-) parallel strahl population.
The presence of Bi-Directional Electrons (BDEs), streaming both parallel and anti-parallel, implies that the field is connected to the Sun at both ends \citep{Palmer1978, Bame1981}.
Hence, the strahl PADs were an important diagnostic tool.
However, this type of data was only available for the Wind and PSP spacecraft in our period of interest.
Therefore, we also calculated the cross helicity, $\sigma_{C}$, of the solar wind in 30 min intervals \citep{Bruno2013, Stansby2018b}.
The magnitude of $\sigma_{C}$ indicates the degree at which there are unidirectional Alfv\'enic fluctuations within a given period, where $|\sigma_{C}| \leq 1$.
The sign of $\sigma_{C}$ indicates the direction of travel for the fluctuations with respect to the magnetic field, with negative (positive) values indicating outward (inward) polarity.
Since Alfv\'enic fluctuations dominantly travel away from the Sun in the plasma frame, $\sigma_{C}$ can be used as a proxy for magnetic polarity.

After assessing the connectivity of a sector boundary, we applied Minimum Variance Analysis (MVA), taking the orientation of the HCS as the plane normal to the minimum variance direction.
Similar to \citet{Burton1994}, we consider the whole sector boundary period, rather than analysing individual directional discontinuities, since \citet{Klein1980} found that these were not a reliable indicator of the overall sector boundary orientation.
For each sector boundary, we applied MVA to a window of duration, T, that was shifted across the event in 10 minute steps.
To ensure the quality of MVA over a given window, it was required that the ratio of the intermediate to minimum eigenvalues, $\lambda_{2}/\lambda_{3} \geq 5$, along with the value of $|B_{n}| / |\vec{B}| < 0.2$ \citep{Jones2000, Kilpua2017a}, where $B_{n}$ is the magnetic field component along the minimum variance direction.
We used the results from the longest duration window with the largest value of $\lambda_{2}/\lambda_{3}$, which can be found in Table \ref{table:events}.

In an attempt to distinguish between transient and co-rotating structures, we have compiled a catalogue of events during our period of interest (Table \ref{table:events}).
To make referencing events more straightforward, we have assigned each event a unique identifier, for example `SO\_2005XN'.
 
\section{Results and discussion}\label{sec:results}
\subsection{HCS structure}\label{sec:results:hcs}

\begin{figure*}[ht!]
\centering
\includegraphics[width=\textwidth]{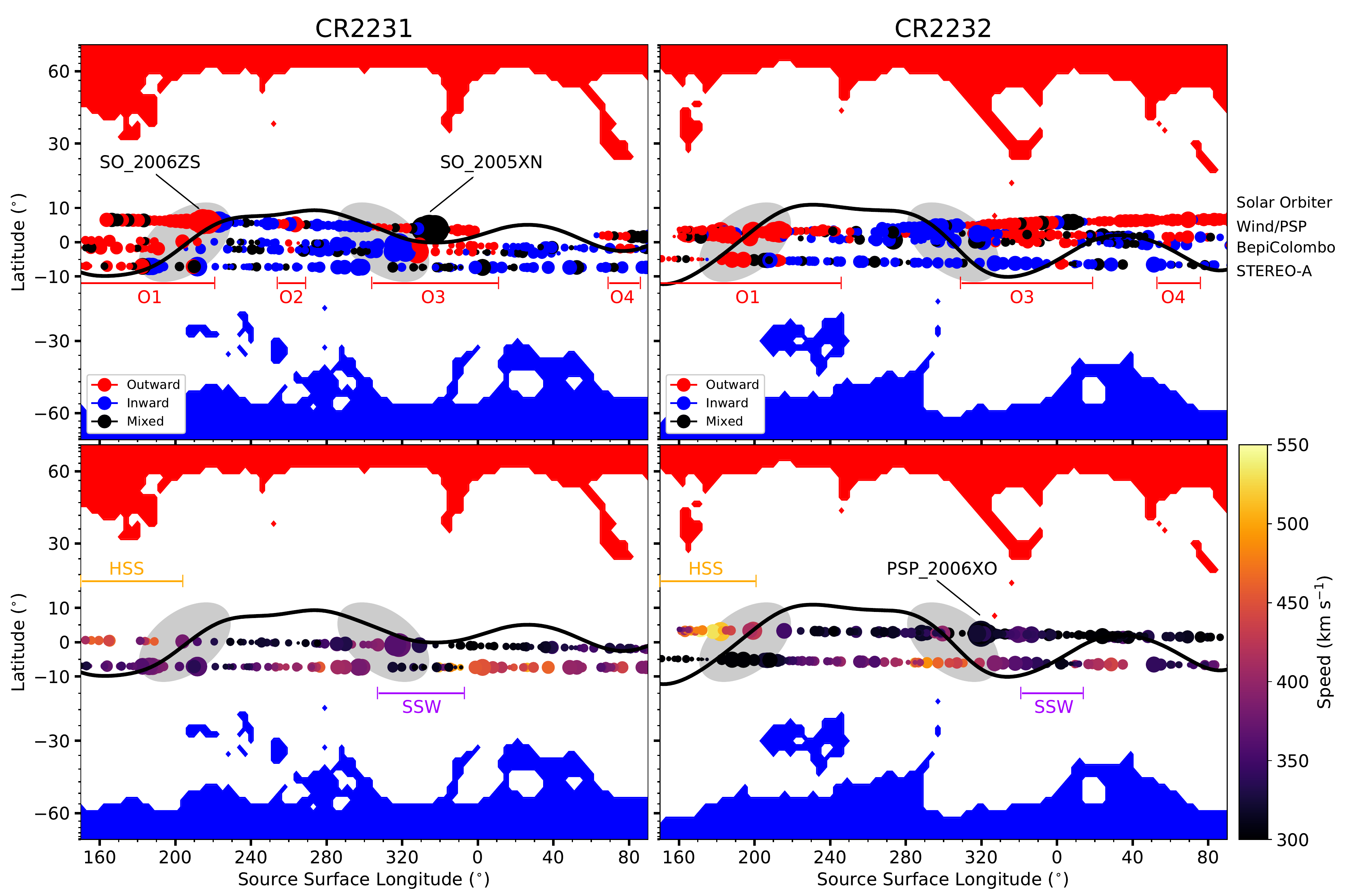}
\caption{Each scatter point represents a 6-hour average of the in situ parameters from a single spacecraft, where the size of the scatter point is proportional to $(|\vec{B}| \times R_{SC})^{2}$ to accentuate changes in $|\vec{B}|$.
The top panels show the polarity measured by the different spacecraft for the two solar rotations, as outlined in Section \ref{sec:method}, with the order of the spacecraft trails next to the top right panel.
In these plots, each spacecraft travelled from right to left, where measurements of the same longitude were made by all the spacecraft within a few days, therefore minimising temporal effects.
Open field lines are shown as the colourmap, as calculated from a PFSS model using an ADAPT magnetogram from the 1st June and 1st July for rotation CR2231 and CR2232, respectively.
We ensured the output of the PFSS model was stable by considering the predicted HCS shape at daily intervals throughout the periods of interest, with the estimated HCS position shown as the solid black line, which broadly matches the shape of the in situ observations.
The bottom left panel shows the solar wind speed from Wind (above) and STEREO-A (below), with the bottom right panel also displaying the data from PSP at the same latitude as Wind.
These maps, along with the open field lines from PFSS, show that the CH structure was stable over the two rotations with a CIRs shown as highlighted grey regions at $\sim200^{\circ}$ and $\sim320^{\circ}$ longitude.
This stable structure is also reflected in the polarity measurements, with outward polarity dips in the HCS outlined with the horizontal red lines.
Transient structures are labelled with the according identification.
}\label{fig:B_v_map}
\end{figure*}

\begin{figure*}[ht!]
\centering
\includegraphics[width=\textwidth]{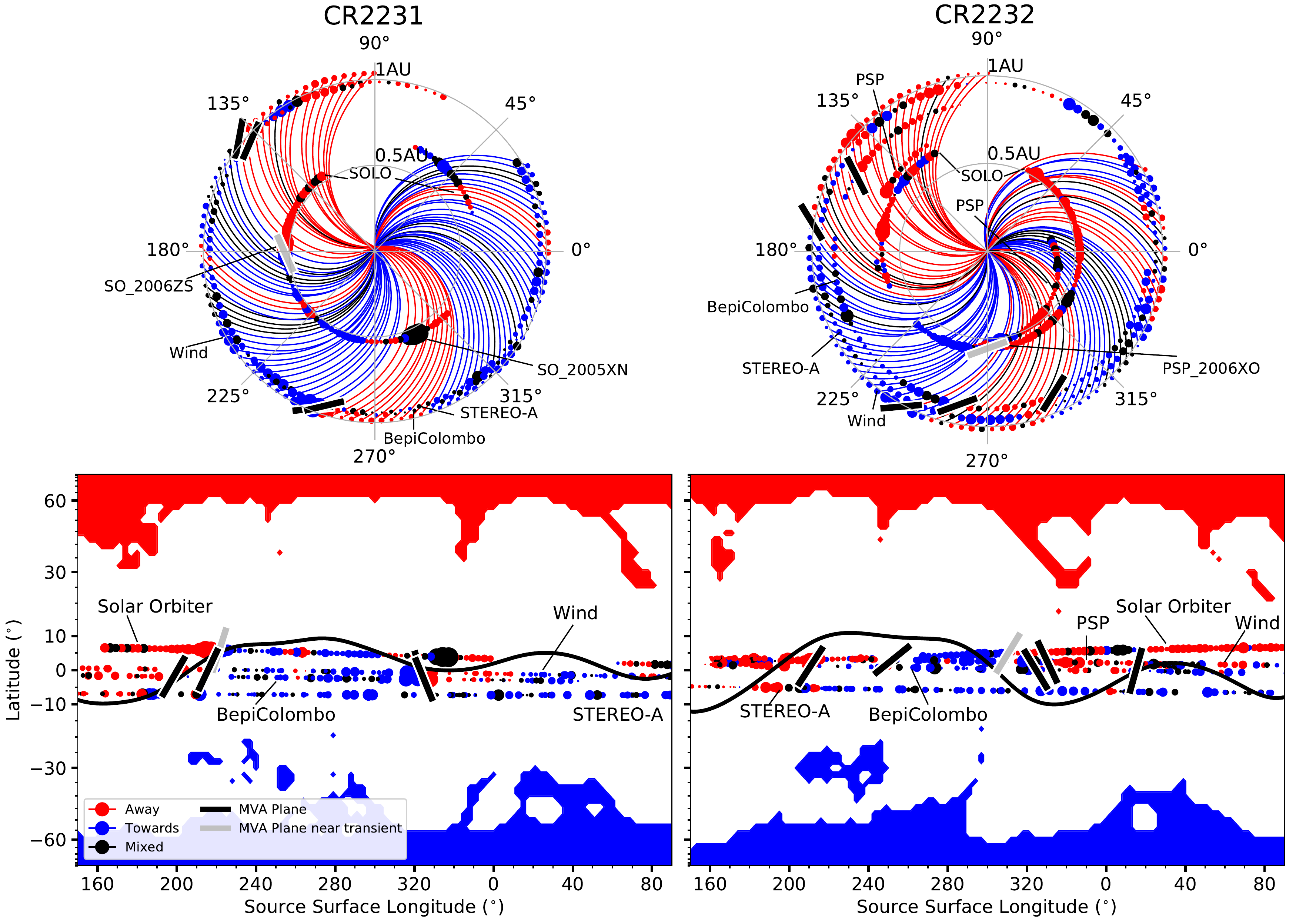}
\caption{Top panels show the in situ measurements in the IAU\_SUN frame, where the angle of the HCS found from MVA is compared to the Parker spiral.
The angles surrounding are of Carrington longitude, which is not the source surface longitude, as plotted in the bottom panels.
The source surface longitude can be estimated by following a Parker spiral line which is plotted based on measurements from Wind.
These angles are generally parallel to the local Parker spiral direction, with the boundaries associated with transients shown as grey.
These plots also give context to where the different spacecraft were during these two rotations, with BepiColombo progressing from $0.99$ AU to $0.85$ AU.
The bottom panels show the source surface as seen in Fig \ref{fig:B_v_map}, with the in situ orientation from different spacecraft overlain, with the values in Table \ref{table:events}.
Although the orientations generally match the direction expected from the HCS, they are much steeper.
}\label{fig:orientation}
\end{figure*}

We applied the ballistic mapping technique, described in Section \ref{sec:method}, to the constellation of spacecraft outlined in Table \ref{table:data} to produce the polarity and speed maps seen in Fig. \ref{fig:B_v_map}.
This demonstrates a clear variation in magnetic polarity with latitude, which reveals that the HCS structure was remarkably flat across these two solar rotations (within $\pm 10^{\circ}$ latitude).
While such a technique has been applied to spacecraft data before \citep{Schwenn1978, Villante1979, burlaga1981}, this constellation provided an unprecedented level of detail in latitude, which was able to resolve several dips in the HCS (regions O1 to O4).
The polarity structure was stable across the two solar rotations, supporting the idea of a stable coronal structure, and therefore HCS shape, which was further evidenced by the similarity in the solar wind speed distribution.
This can be seen in the bottom axes of Fig. \ref{fig:B_v_map} where a high speed stream (HSS) was observed by Wind and PSP at longitudes $<200^{\circ}$ which, along with other speed gradients, led to the formation of several CIRs highlighted in grey.

The estimated HCS position from a PFSS model (black line), generally matches the shape of the HCS from the in situ measurements, and produces similar open field line distributions across the two solar rotations (colourmap in Fig. \ref{fig:B_v_map}).
While we see a consistent shape between the PFSS model and in situ measurements, we do not attempt a more detailed comparison as this would involve more careful application of PFSS, that may require adjusting the source surface height \citep{Badman2020a, Panasenco2020, Kruse2021}, or using more complex models \citep{Odstrcil2003, Jian2015, Pomoell2018}.
However, we do note that the number of measurements over this range of latitudes could be used to better constrain the polarity and solar wind speed predicted by solar wind models.
This would be most relevant at solar minimum, since Fig \ref{fig:B_v_map} shows that a few degrees of latitude can drastically change the solar wind conditions experienced by a spacecraft, which agrees with the latitude scale size found by previous studies \citep{Schwenn1978, Owens2020a}.

The local orientation of the sector boundaries can be seen in Fig. \ref{fig:orientation}, where the Parker spiral lines are drawn from the position of the Wind spacecraft, using the measured solar wind speed.
In general, the plane measured from MVA was aligned along the Parker spiral direction in the R-T plane (top panels), and the orientation out the R-T plane (bottom panels) matched the sense of the inclination from the PFSS model.
This was also consistent with the shape traced out by several spacecraft crossings of the same sector boundary at different latitudes, best demonstrated with region O1 in CR2231.
These results support previous studies (at solar maximum) that found the PFSS model agreed with the in situ orientation, in the absence of any transient structures \citep{Klein1980, Burton1994, Peng2017}.
Unlike these studies, we measured the local orientation to be steeper than the relatively flat HCS seen with in situ observations and the PFSS model.
We note that this is most likely due to the presence of stream interactions at these boundaries, which steepen with distance from the Sun \citep{Pizzo1991}.
Therefore, these observations could represent the start of HCS distortion, that is known to be significant further from the Sun \citep[$> 2$ AU, ][]{Pizzo1994, Lee2000, Riley2002}.
Although we have also estimated the orientation of boundaries at $0.5$ AU, they cannot be used as evidence of this steepening argument as they are in close proximity to transient events (shown in grey in Fig. \ref{fig:orientation}) that will be discussed in the Section \ref{sec:results:transient}.

\subsection{Co-rotating Interaction Regions}\label{sec:results:cir}

\begin{figure}[ht!]
\centering
\includegraphics[width=\hsize]{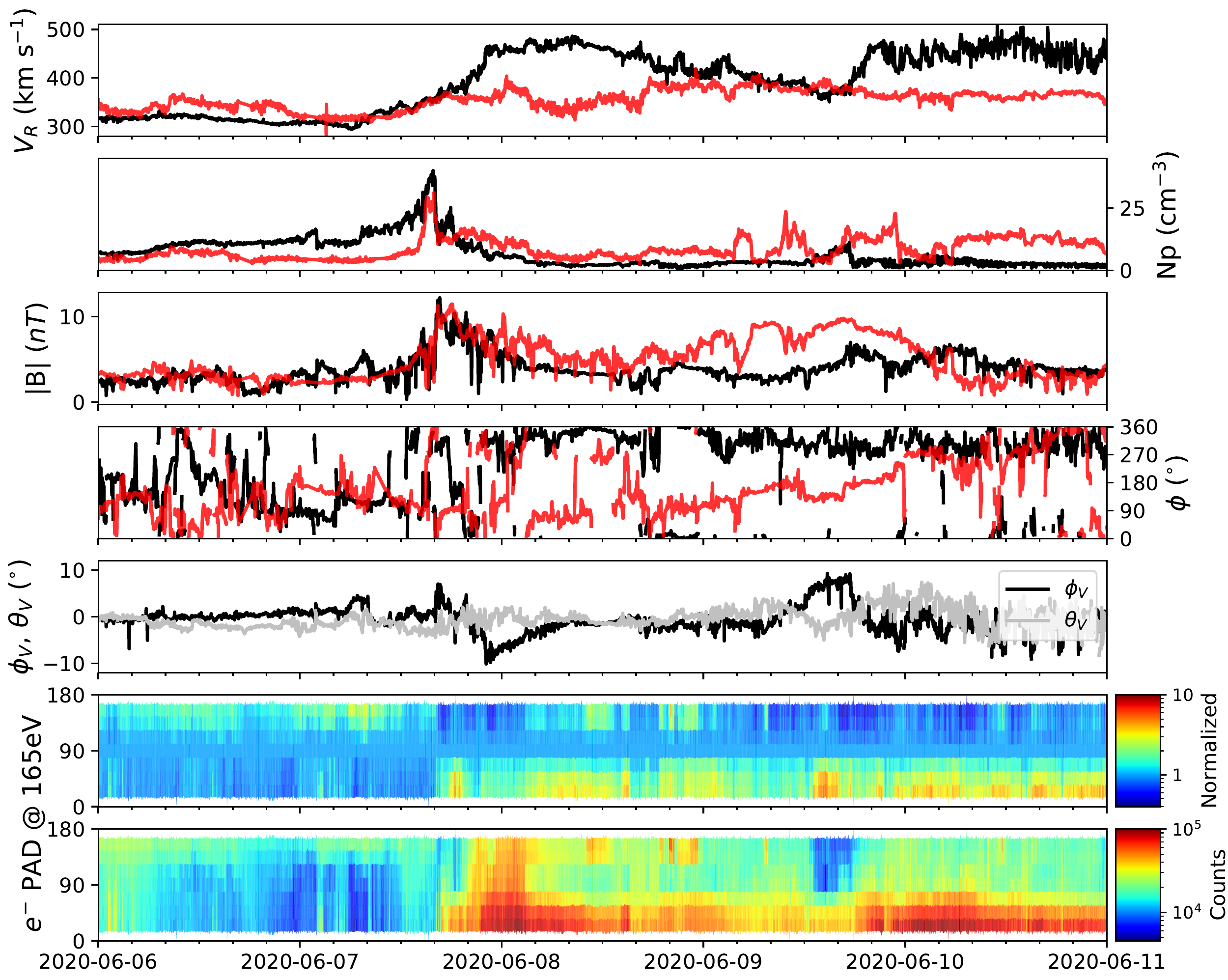}
\caption{The CIR observed by Wind and STEREO-A which represents the region O1 in Fig. \ref{fig:B_v_map}. Here STEREO-A measurements (red), from the same source surface longitude, have been time shifted by $\sim 6$ days to line up with the increase in |B|. Wind observes a clear increase in solar wind speed from $300$ km s$^{-1}$ to $465$ km s$^{-1}$, along with a distinctive east-west (positive-negative) deflection in the $\phi_{V}$ signature. Both spacecraft observe a similar increase in $|\vec{B}|$ and $N_{p}$, but STEREO-A only measures an increase of $320$ km s$^{-1}$ to $350$ km s$^{-1}$.
}\label{fig:wind_200}
\end{figure}

Instead of applying statistics to a large number of CIRs, this spacecraft constellation allows the study of how individual CIRs vary with latitude.
One such example is the CIR around $200^{\circ}$ longitude in CR2231, which was measured by four spacecraft, with the time series of Wind and STEREO-A data being shown in Fig. \ref{fig:wind_200}.
Wind observed the solar wind speed to increase from $300$ km s$^{-1}$ to $465$ km s$^{-1}$, along with the typical flow deflections expected from a CIR (Fig. \ref{fig:wind_200} panel 5).
This HSS originated from the northern polar coronal hole (CH).
Therefore, SO, at a higher latitude than Wind, was connected deeper into the CH, which was supported by the presence of large amplitude Alfv\'enic fluctuations \citep{Belcher1971}.
STEREO-A, at $-7^{\circ}$ latitude, only measured an increase in solar wind speed from $320$ km s$^{-1}$ to $350$ km s$^{-1}$, but recorded a similar $|B|$ profile and increase in proton density.
This type of compression, without a clear HSS at similar latitude, has been observed with earlier spacecraft \citep{Burlaga1983, Schwenn1990} and is present in the same CIR on the next solar rotation (CR2232 of Fig. \ref{fig:B_v_map}).

These measurements can be explained by considering the tilt of this particular CIR, as seen in Fig \ref{fig:orientation}.
This introduced a southern component to the direction of the forward propagating pressure wave at the leading edge of the CIR, which allowed the HSS to compress slow solar wind beyond its own latitudinal extent.
Further from the Sun, this forward propagating compression wave will likely steepen into a forward shock, which have been shown to propagate perpendicular to the stream interaction boundary with the Ulysses spacecraft \citep{Gosling1999} and numerical models \citep{Riley2012}.
Although this forward shock has not yet developed in these measurements (at 1 AU), we have shown that the solar wind is still affected on scales of a few degrees in latitude by the CIR tilt.

This corroborates the findings from the twin Helios spacecraft \citep{Schwenn1978}, demonstrating the importance of latitude when investigating CIRs, which is also evident in the other CIR $\sim300^{\circ}$ longitude in Fig. \ref{fig:B_v_map}.
Here STEREO-A measured a gradient solar wind speed, along with an increase in $|\vec{B}|$ and proton density.
However, unlike typical CIR observations, STEREO-A did not measure a change in the magnetic polarity.
Therefore, these measurements represent the lower extent in latitude of a CIR without the presence of the HCS, which may alter how the CIR develops.
We verified the presence of typical CIR flow deflections in STEREO-A, although the measurements were not reliable enough to carry out a detailed analysis on the CIR dynamics.
While we were limited by the lack of plasma and electron measurements from some spacecraft, we have demonstrated how such a multi-spacecraft study can isolate changes in latitude for an individual CIR, which may be important for space weather prediction.

\subsection{Identifying co-rotating structure}\label{sec:results:iden}

The spread of spacecraft latitude can also provide contextual information, allowing for identification of solar wind structures that would otherwise not be possible.
One such example is the small scale dip in the HCS labelled as O2 in the top left panel of Fig. \ref{fig:B_v_map}.
This outward polarity region was observed by SO, Wind and BepiColombo over a range of $-2.3^{\circ}$ to $5.4^{\circ}$ latitude, and spanned $\sim15^{\circ}$ longitude in SO.
The initial change from inward to outward magnetic field was measured by SO at 16:00 on 3 June 2020 at $0.5$ AU, which was followed by BepiColombo $\sim2$ hours later at 1 AU, and at Wind a further $\sim13$ hours after.
Since BepiColombo and Wind were both orbiting at 1 AU during CR2231 (as seen in Fig \ref{fig:orientation}), their separation in longitude meant that BepiColombo should see a co-rotating structure roughly $12$ hours prior to Wind.
This, along with the short time interval between BepiColombo and SO, suggests that this field reversal was a co-rotating structure, most likely associated with a ripple in the HCS of similar size to those proposed by \citet{Gosling1981} and \citet{Villante1979}.
Wind observations of the electron strahl confirmed that this field reversal was related to a sector boundary, with the longer period in SO implying that this was a fine scale ripple in the HCS towards lower latitudes.

There was extra complexity in the SO measurements, with exotic wave activity and an additional inward pointing magnetic field region in the otherwise outward polarity region.
This could be due to magnetic field draping and pick-ions related to comet ATLAS, which SO was predicted to interact with \citep{Jones2020} and is studied in more detail by \citet{Matteini2021}.
STEREO-A measurements at $-7^{\circ}$ latitude showed no reversal in polarity, and observed the trailing edge of a HSS, unlike Wind measurements of $315$ km s$^{-1}$, which implies that this dip does not extend much further south than BepiColombo.
This also provided an upper limit of $5^{\circ}$ latitude for the thickness of the streamer belt, which is consistent with earlier observations \citep{Richardson1997, Chen2021}.
Therefore, mapping the speed of the solar wind can further constrain the location of the HCS, rather than relying solely on the magnetic field polarity.

Another example of this is the outward polarity region O3 which, like O2, did not exhibit a magnetic field reversal in STEREO-A in CR2231.
However, in this case STEREO-A did measure slow solar wind (labelled SSW in Fig \ref{fig:B_v_map}) which confirms that O3 extended to lower latitudes than O2.
Using the $5^{\circ}$ estimated width of the streamer belt, we conclude that STEREO-A was likely at the edge of the streamer belt.
Such a classification could be useful for studying specific situations in the solar wind, like the formation of the Kelvin-Helmholtz instability.

Although the spacecraft used in this paper were not a coordinated effort to create a multi-spacecraft mission, their wide variety of orbits can be used to isolate a particular aspect of solar wind evolution: either time, latitude or distance from the Sun.
In the future, these opportunities will be able to exploit different aspects of each spacecraft's orbit, since SO will become more inclined to the ecliptic plane (by around $30^{\circ}$) and PSP will measure the solar wind at progressively closer distances to the Sun.
 
\subsection{Transients}\label{sec:results:transient}

\begin{figure*}[!ht]
\centering
\includegraphics[width=\textwidth]{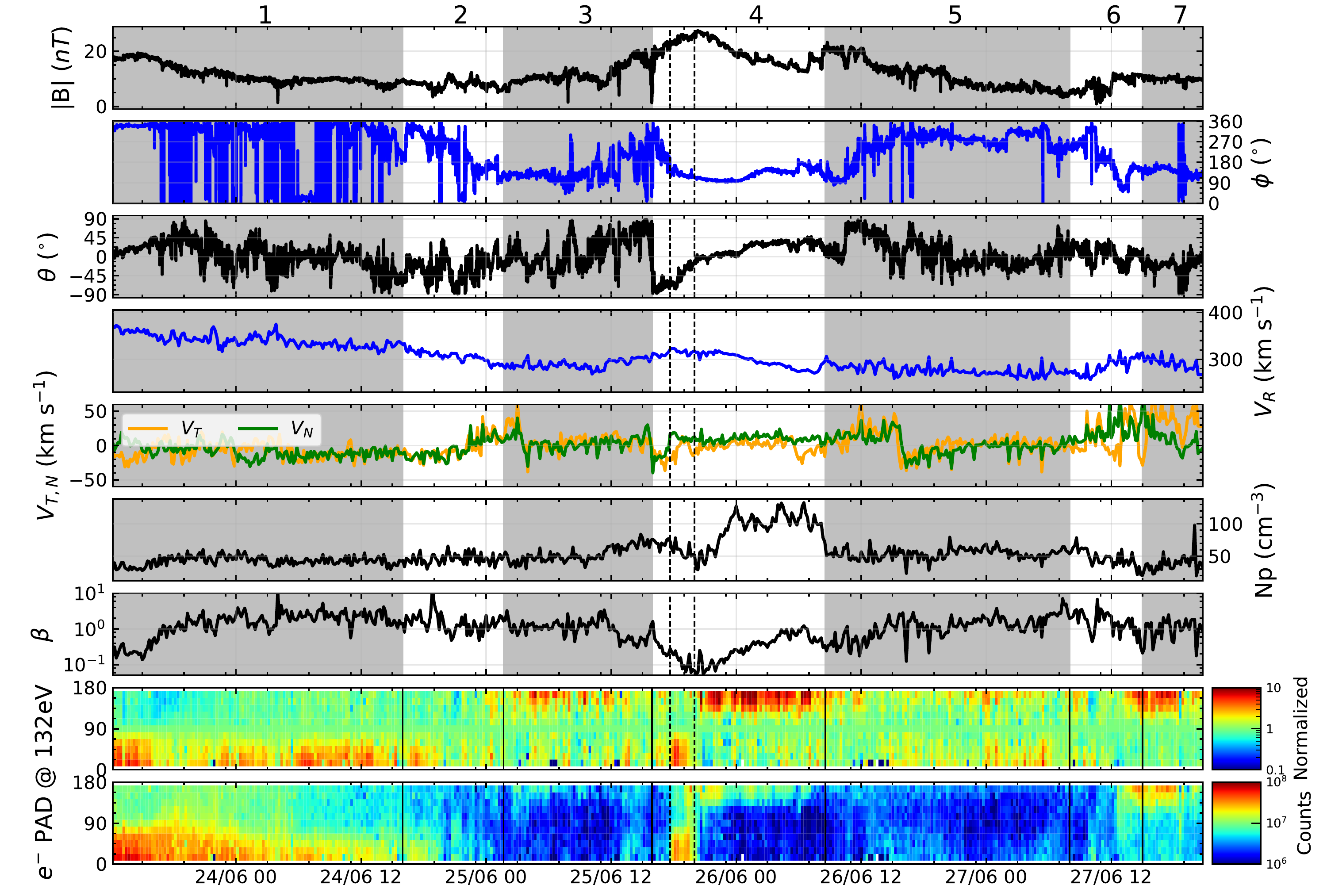}
\caption{Timeseries of a sector boundary observed by PSP, with ambient solar wind conditions seen in regions 1 and 7, and the transient event (PSP\_2006XO) appearing in region 4.
This was classified as an ICME due to the decreasing radial speed, reduction in $\beta$ and smooth magnetic field rotation.
The plasma parameters have been smoothed over 30 minutes.
The penultimate panel shows the electron strahl PAD, normalised to its value at $90^{\circ}$.
The electron signature, in the bottom two panels, weakens in regions 3 and 5 which could be due to scattering processes.
}\label{fig:psp}
\end{figure*}

By comparing measurements from multiple spacecraft, we were also able to classify several transient structures, which are labelled as such in Table \ref{table:events}.
The most complex of these events was SO\_2006ZS, labelled in Figs. \ref{fig:B_v_map} and \ref{fig:orientation}, which was likely the combination of two CMEs, a CIR and a HCS crossing, which is investigated in detail by \citet{Telloni2021}.

A flux rope like structure was observed by SO on 29 May 2020 (SO\_2005XN) at $\sim0.5$ AU (Fig \ref{fig:solo}).
This structure displayed a smooth rotation in the magnetic field with a minimium variance direction along the radial.
A similar structure was observed on the following solar rotation at $\sim0.5$ AU by PSP on 26 June 2020 (PSP\_2006XO), as shown in Fig \ref{fig:psp}.
Again, this was a flux rope like structure with an increase in |\vec{B}|, $\theta$ ranging from $-80^{\circ}$ to $45^{\circ}$ and was confined to a plane with a normal direction along \vec{R}.
Like SO\_2005XO, this was seen at around $280^{\circ}$ Carrington longitude, which may suggest that these were co-rotating structures.
However, we conclude these are interplanetary CMEs (ICMEs) due to absence of these structures in other spacecraft, low plasma $\beta$ and the smoothly decreasing radial speed in region 4 of Fig \ref{fig:psp}, which is indicative of ICME expansion \citep{Zurbuchen2006, Richardson2010a}.

Therefore, these measurements could represent the same type of ICME from the same source region and similar distances from the Sun.
While we have identified possible CMEs in remote sensing observations\footnote{26 May 2020 00:39 for SO\_2005XN and 22 June 2020 15:09 for PSP\_2006XO in the DONKI catalogue \url{https://kauai.ccmc.gsfc.nasa.gov/DONKI/}}, we leave modelling of these events for a future study and instead focus on the in situ characteristics.

The ICME, in region 4 of Fig \ref{fig:psp}, appears in conjunction with the HCS crossing, which transitions from outward polarity with parallel strahl electrons in region 1, to inward polarity and anti-parallel strahl in region 7.
Similar ICMEs have been observed previously, and have been interpreted as magnetic clouds that are part of the HCS, rather than pushing it aside or draping \citep{Crooker1993, Crooker1996, forsyth1997}.

Within region 4, there was a period of parallel electrons (bounded by dashed vertical lines) with an inward magnetic field, implying these field lines were folded by the ICME, which could explain why the strahl direction changes when $\theta$ changes sign.
The flux rope itself (after the latter dashed line) had anti-parallel, rather than counter-streaming electrons, implying it was connected to the Sun at one end \citep{Crooker2008}.
This indicates magnetic reconnection has occurred, either during interchange reconnection as it escaped from the Sun \citep{Crooker1993, Gosling1995} or as it propagated through the solar wind \citep{McComas1994}.
Following the ICME, the magnetic field returned to outward magnetic polarity in region 5, with evidence of weak counter-streaming electrons.
This could imply the field lines were connected at both ends to the Sun, although it was more likely due to reflection or focusing of electrons at some boundary in this complex magnetic structure \citep{Gosling2001}.

While Fig \ref{fig:psp} displays an ICME within a sector boundary, this same HCS crossing without an ICME was observed by Wind and BepiColombo at a similar time, which represented an earlier configuration of the solar wind from the same source region.
Wind observed an increase in |\vec{B}| and solar wind speed, indicating a weak CIR (WIND\_2006JJ). The minimum variance normal was along ($34^{\circ} \ \phi$, $26^{\circ} \ \theta$) and ($33^{\circ} \ \phi$, $32^{\circ} \ \theta$) for Wind and BepiColombo respectively.
In contrast, SO experienced this HCS crossing $\sim5$ days after the passage of the ICME in PSP, representing the same source region at a later date and observed the sector boundary plane with a normal along ($20^{\circ} \ \phi$, $-32^{\circ} \ \theta$).
We propose that this was due to the HCS still reforming after the eruption of a CME that was previously observed by PSP.

While it is well known that the HCS reforms after the passage of a CME \citep{Zhao1996, Blanco2011}, it is unclear over what timescale this process takes place.
If these SO measurements do indeed represent a disrupted HCS, then we argue that this puts a lower limit of $\sim5$ days on this reformation.
Similar timescales have been found ($\sim 3$ to $6$ days) in numerical studies that simulated the passage of a CME through background solar wind \citep{Temmer2017, Desai2020}.
However, the lack of plasma and electron measurements at SO make this claim purely speculative, although it does demonstrate the applicability of multi-spacecraft studies on transient solar wind structures.
 
\section{Conclusions}\label{sec:conclusions}
After the recent launches of Parker Solar Probe and Solar Orbiter, there are an unprecedented number of spacecraft simultaneously measuring the solar wind.
During the period of May to July 2020 these spacecraft, along with BepiColombo, Wind and STEREO-A, provided the opportunity to study the solar wind at a range of latitudes.
By mapping the magnetic polarity measured by each spacecraft, we have demonstrated how the structure and position of the HCS can be investigated, finding that:
\begin{itemize}
  \item The solar wind measured by a spacecraft at solar minimum depended largely on the latitude of the observation.
  \item The HCS was remarkably flat over two solar rotations (CR2231 and CR2232) between May and July 2020, meaning we were able to resolve fine scale ripples in the HCS down to scales of several degrees in latitude.
  \item The location of the HCS was further constrained by examining the solar wind speed at each spacecraft, as this could reveal times when a spacecraft was within the slow solar wind surrounding the HCS without changing polarity.
  \item A PFSS model captured the general shape of the HCS, and agreed with the locally measured sector boundary orientations, but were much steeper. However, due to the presence of transient events we could not observe how the HCS steepened between $0.5$ and $1$ AU.
  \item A CIR was measured at four different latitudes, which showed that compressed slow solar wind was observed by STEREO-A even in the absence of an accompanying high speed stream. This is evidence for the compression region propagating in latitude.
  \item Another CIR was observed by STEREO-A without any change in magnetic polarity, highlighting how important latitude is to the CIR structure.
  \item We could identify several transient structures around sector boundaries, that we classified as ICMEs. By observing a HCS crossing before, during and after an ICME interaction we found evidence of distortion that lasted at least $5$ days.
\end{itemize}

These results highlight that at solar minimum the solar wind varies on scales of just a few degrees in latitude, which can drastically alter the solar wind conditions measured by different spacecraft.
Therefore, this combination of spacecraft can be used to constrain solar wind models in latitude, which could improve space weather predictions.

We also note that by comparing measurements across these spacecraft, we can give context to solar wind measurements, that would not otherwise be possible.
This could open up new research opportunities since one can be more confident of where a spacecraft is in relation to large scale structures - i.e. crossing the bottom of a CIR, or skimming the HCS.

It is important to note that this collection of spacecraft were not intended to act as a multi-spacecraft mission.
Therefore, each spacecraft has its own unique orbital characteristics, with PSP going closer to the Sun; BepiColombo travelling to Mercury and SO increasing the inclination of its orbit.
This means that the configuration of the spacecraft will evolve with time, which will continue to provide unique avenues for future heliospheric research.

\begin{acknowledgements}
RL was supported by an Imperial College President's Scholarship, TSH and JPE by STFC ST/S000364/1, TW by ST/N504336/1, LDW by ST/S000364/1, CJO by ST/5000240/1. SDB acknowledges the support of the Leverhulme Trust Visiting Professor program. The SWEAP and FIELDS teams acknowledge support from NASA contract NNN06AA01C. The Solar Orbiter magnetometer was funded by the UK Space Agency (grant ST/T001062/1). The Solar Orbiter Solar Wind Analyser (SWA) PAS were designed, created, and are operated under funding provided in numerous contracts from the UK Space Agency (UKSA), the UK Science and Technology Facilities Council (STFC), the Centre National d’Etudes Spatiales (CNES, France), the Centre National de la Recherche Scientifique (CNRS, France), and the Czech contribution to the ESA PRODEX programme. . Solar Orbiter data are available from the Solar Orbiter Archive at http://soar.esac.esa.int/soar/.  This work has made use of the open source and free community-developed space physics packages HelioPy \citep{heliopy2020}, pfsspy \citep{pfsspy} and SpiceyPy \citep{spiceypy2020}. Solar Orbiter is a space mission of international collaboration between ESA and NASA, operated by ESA.
\end{acknowledgements}

\bibliographystyle{aa} \bibliography{Orb_spec_issue.bbl}

\section*{Appendix A: Additional Material}
\renewcommand{\thefigure}{A.\arabic{figure}}
\setcounter{figure}{0}
\renewcommand{\thetable}{A.\arabic{table}}
\setcounter{table}{0}

\begin{table}[h!]
\resizebox{\hsize}{!}{\begin{tabular}{@{}c|c|l|c|c|c|c|l|c|l@{}}
\toprule
Event                 & Date                  & Type & $\Phi$                & $\Theta$              & $R$                   & $\Phi_{SS}$           & MVA Start        & MVA ($\phi$, $\theta$)   & $\lambda_{2}/\lambda_{3}$ \\
                      &                       &      & ($^{\circ}$)          & ($^{\circ}$)          & (AU)                  & ($^{\circ}$)          & End              & ($^{\circ}$, $^{\circ}$) &                           \\ \midrule
SO\_2005XN            & 29/05/2020 00:00      & ICME & 290                   & 3.9                   & 0.56                  & 330                   & -                & -                        &                           \\ \midrule
BC\_2005OW            & 29/05/2020 15:30      & HCS  & 252                   & -2.7                  & 0.97                  & 325                   & 29/05/2020 08:20 & (31, 23)                 & 18                        \\
\multicolumn{1}{l|}{} & \multicolumn{1}{l|}{} & CIR  & \multicolumn{1}{l|}{} & \multicolumn{1}{l|}{} & \multicolumn{1}{l|}{} & \multicolumn{1}{l|}{} & 30.05/2020 08:20 & \multicolumn{1}{l|}{}    &                           \\ \midrule
Wind\_2005AW          & 30/05/2020 04:18      & HCS  & 250                   & -0.9                  & 1.01                  & 320                   & 29/05/2020 15:51 & (27,18)                  & 13                        \\
\multicolumn{1}{l|}{} & \multicolumn{1}{l|}{} & CIR  & \multicolumn{1}{l|}{} & \multicolumn{1}{l|}{} & \multicolumn{1}{l|}{} & \multicolumn{1}{l|}{} & 30/05/2020 15:51 & \multicolumn{1}{l|}{}    &                           \\ \midrule
BC\_2006LP            & 06/06/2020 18:30      & HCS  & 140                   & -1.9                  & 0.95                  & 198                   & 07/06/2020 01:50 & (17,-31)                 & 5.2                       \\
\multicolumn{1}{l|}{} & \multicolumn{1}{l|}{} & CIR  & \multicolumn{1}{l|}{} & \multicolumn{1}{l|}{} & \multicolumn{1}{l|}{} & \multicolumn{1}{l|}{} & 07/06/2020 13:50 & \multicolumn{1}{l|}{}    &                           \\ \midrule
Wind\_2006HR          & 07/06/2020 12:00      & HCS  & 140                   & 0.1                   & 1.01                  & 216                   & 07/06/2020 03:48 & (28, -26)                & 5.9                       \\
\multicolumn{1}{l|}{} & \multicolumn{1}{l|}{} & CIR  & \multicolumn{1}{l|}{} & \multicolumn{1}{l|}{} & \multicolumn{1}{l|}{} & \multicolumn{1}{l|}{} & 08/06/2020 03:48 & \multicolumn{1}{l|}{}    &                           \\ \midrule
SO\_2006ZS            & 07/06/2020 18:00      & HCS  & 184                   & 5.9                   & 0.53                  & 220                   & 07/06/2020 16:10 & (20, -18)                & 11                        \\
\multicolumn{1}{l|}{} & \multicolumn{1}{l|}{} & CIR  & \multicolumn{1}{l|}{} & \multicolumn{1}{l|}{} & \multicolumn{1}{l|}{} & \multicolumn{1}{l|}{} & 08/06/2020 16:10 & \multicolumn{1}{l|}{}    &                           \\
\multicolumn{1}{l|}{} & \multicolumn{1}{l|}{} & ICMEs & \multicolumn{1}{l|}{} & \multicolumn{1}{l|}{} & \multicolumn{1}{l|}{} & \multicolumn{1}{l|}{} &                  & \multicolumn{1}{l|}{}    &                           \\ \midrule
BC\_2006YZ            & 22/06/2020 21:00      & HCS  & 260                   & -0.2                  & 0.9                   & 375                   & 22/06/2020 14:30 & (31, -16)                & 10.5                      \\
\multicolumn{1}{l|}{} & \multicolumn{1}{l|}{} &      & \multicolumn{1}{l|}{} & \multicolumn{1}{l|}{} & \multicolumn{1}{l|}{} & \multicolumn{1}{l|}{} & 23/06/2020 02:30 & \multicolumn{1}{l|}{}    &                           \\ \midrule
BC\_2006YU            & 25/06/2020 14:00      & HCS  & 295                   & 0.1                   & 0.89                  & 325                   & 25/06/2020 10:00 & (33, 32)                 & 17.5                      \\
\multicolumn{1}{l|}{} & \multicolumn{1}{l|}{} & CIR  & \multicolumn{1}{l|}{} & \multicolumn{1}{l|}{} & \multicolumn{1}{l|}{} & \multicolumn{1}{l|}{} & 25/06/2020 22:00 & \multicolumn{1}{l|}{}    &                           \\ \midrule
PSP\_2006XO           & 26/06/2020 00:00      & ICME & 280                   & 2.4                   & 0.52                  & 316                   & -                & -                        &                           \\ \midrule
Wind\_2006JJ          & 26/06/2020 13:00      & HCS  & 241                   & 2.4                   & 1.02                  & 330                   & 26/06/2020 13:00 & (34, 26)                 & 10.2                      \\
\multicolumn{1}{l|}{} & \multicolumn{1}{l|}{} & CIR  & \multicolumn{1}{l|}{} & \multicolumn{1}{l|}{} & \multicolumn{1}{l|}{} & \multicolumn{1}{l|}{} & 27/06/2020 13:00 & \multicolumn{1}{l|}{}    &                           \\ \midrule
SO\_2007KG            & 02/07/2020 04:00      & HCS  & 270                   & 5                     & 0.56                  & 310                   & 01/07/2020 22:00 & (20, -32)                & 6.1                       \\
\multicolumn{1}{l|}{} & \multicolumn{1}{l|}{} &      & \multicolumn{1}{l|}{} & \multicolumn{1}{l|}{} & \multicolumn{1}{l|}{} & \multicolumn{1}{l|}{} & 02/07/2020 10:00 & \multicolumn{1}{l|}{}    &                           \\ \midrule
Wind\_2007BU          & 02/07/2020 11:00      & HCS  & 170                   & 3                     & 1.02                  & 250                   & 02/07/2020 14:05 & (41, -52)                & 7.2                       \\
\multicolumn{1}{l|}{} & \multicolumn{1}{l|}{} & CIR  & \multicolumn{1}{l|}{} & \multicolumn{1}{l|}{} & \multicolumn{1}{l|}{} & \multicolumn{1}{l|}{} & 03/07/2020 02:05 & \multicolumn{1}{l|}{}    &                           \\ \midrule
BC\_2007JH            & 03/07/2020 23:00      & HCS  & 150                   & 1.1                   & 0.86                  & 210                   & 03/07/2020 17:50 & (57, -34)                & 6                         \\
\multicolumn{1}{l|}{} & \multicolumn{1}{l|}{} & CIR  & \multicolumn{1}{l|}{} & \multicolumn{1}{l|}{} & \multicolumn{1}{l|}{} & \multicolumn{1}{l|}{} & 03/07/2020 23:50 & \multicolumn{1}{l|}{}    &                           \\ \bottomrule
\end{tabular}}
\caption{List of events with significant MVA results used in this paper. The string before the underscore denotes the spacecraft that made the measurement; the digits represent the two digits year and month with the final two random characters differentiating the events within each month. MVA was used to determine the orientation of the boundary, and is left blank for CMEs.}
\label{table:events}
\end{table}

\begin{figure}[ht!]
\centering
\includegraphics[width=\hsize]{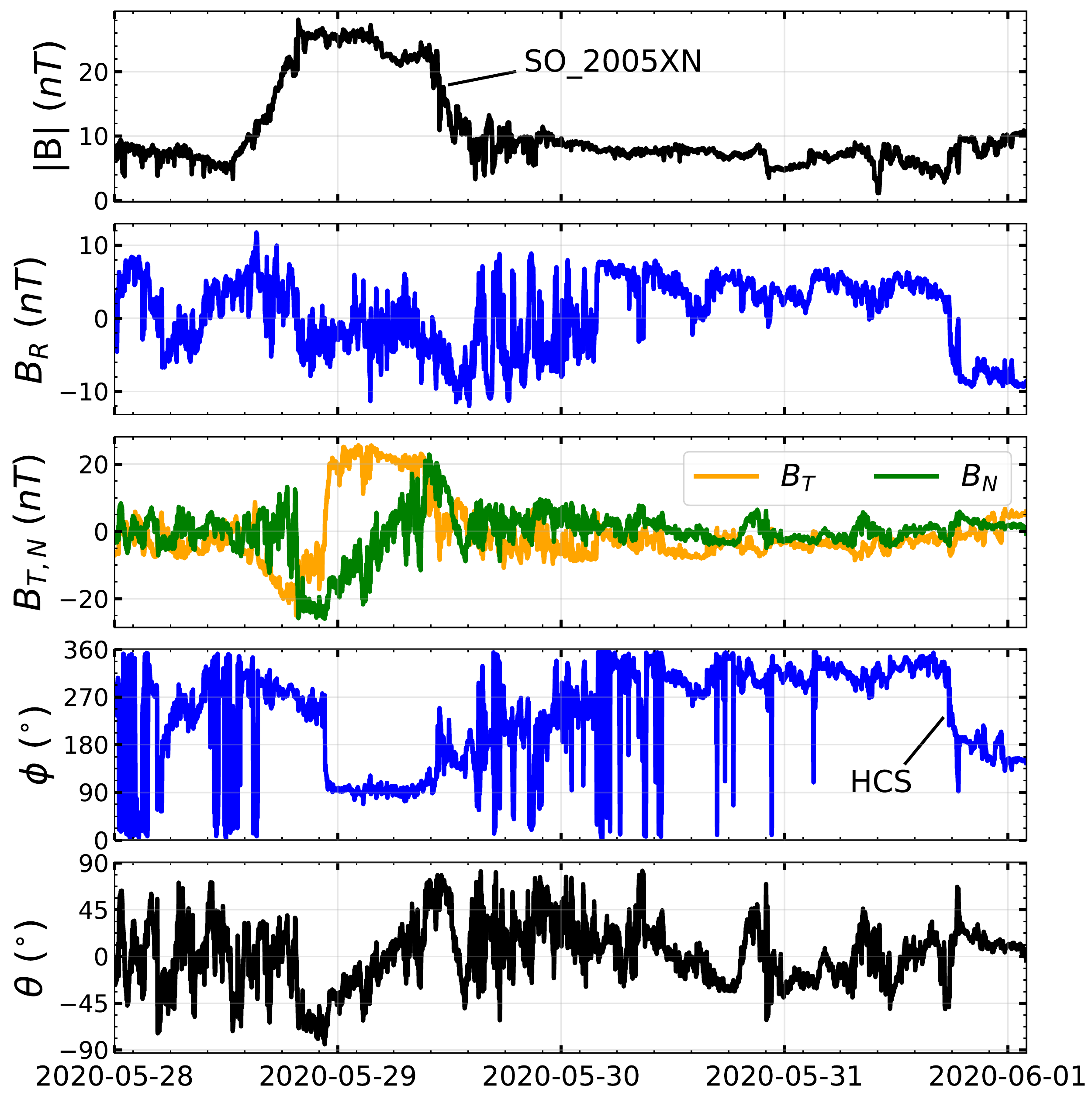}
\caption{Transient feature, SO\_2005XN, seen by SO during CR2231.
There is a smooth rotation in the magnetic field, indicating a flux rope with radial minimum variance direction.
PAS measurements indicate that the sign of $\sigma_{C}$ reverses across the boundary marked as a HCS crossing.
A similar flux rope structure was observed on the next solar rotation by PSP, Fig. \ref{fig:psp}, at the same distance and longitude.
Like the PSP observations, there is a reversal in the field, although this occurs $\sim3$ days after the flux rope in this case.
}\label{fig:solo}
\end{figure}

\end{document}